\documentclass[twocolumn,aps]{revtex4}
\usepackage{color}
\usepackage{graphicx}

\newcommand{\Fe}{Fe$_{4}$}
\newcommand{\FeB}{Fe$_{4}^{\rm B}$}
\newcommand{\FeML}{Fe$_{4}^{\rm ML}$}
\newcommand{\musr}{$\mu$SR}
\newcommand{\lem}{LE-$\mu$SR}

\begin{document}
\title{Proximal magnetometry of monolayers of single molecule magnets on
gold using polarized muons}
\author{Z.~Salman}
\email{zaher.salman1@psi.ch}
\affiliation{Clarendon Laboratory, Department of Physics, Oxford University, Parks Road, Oxford OX1 3PU, UK}
\affiliation{Paul Scherrer Institute, Laboratory for Muon Spin Spectroscopy, CH-5232 Villigen PSI, Switzerland}
\author{S.J.~Blundell}
\affiliation{Clarendon Laboratory, Department of Physics, Oxford University, Parks Road, Oxford OX1 3PU, UK}
\author{S.R.~Giblin}
\affiliation{ISIS Facility, Rutherford Appleton Laboratory, Chilton, Oxfordshire, OX11 0QX, UK}
\author{M.~Mannini}
\affiliation{Dipartimento di Chimica, Universit\`a di Firenze \& INSTM, via della
Lastruccia 3, 50019 Sesto Fiorentino, Italy}
\affiliation{ISTM-CNR, URT Firenze, 50019 Sesto Fiorentino, Italy.}
\author{L.~Margheriti}
\affiliation{Dipartimento di Chimica, Universit\`a di Firenze \& INSTM, via della
Lastruccia 3, 50019 Sesto Fiorentino, Italy}
\author{E.~Morenzoni}
\author{T.~Prokscha}
\author{A.~Suter}
\affiliation{Paul Scherrer Institute, Laboratory for Muon Spin Spectroscopy, CH-5232 Villigen PSI, Switzerland}
\author{A.~Cornia}
\affiliation{Department of Chemistry \& INSTM Research Unit, Universit\`a degli
Studi di Modena e Reggio Emilia, Via G. Campi 183, 41100 Modena, Italy}
\author{R.~Sessoli}
\affiliation{Dipartimento di Chimica, Universit\`a di Firenze \& INSTM, via della
Lastruccia 3, 50019 Sesto Fiorentino, Italy}

\begin{abstract}
  The magnetic properties of a monolayer of \Fe\ single molecule
  magnets grafted onto a Au (111) thin film have been investigated
  using low energy muon spin rotation. The properties of the monolayer
  are compared to bulk \Fe. We find that the magnetic properties in
  the monolayer are consistent with those measured in the bulk,
  strongly indicating that the single molecule magnet nature of \Fe\
  is preserved in a monolayer. However, differences in the temperature
  dependencies point to a small difference in their energy scale. We
  attribute this to a $\sim 60 \%$ increase in the intramolecular
  magnetic interactions in the monolayer.
\end{abstract}
\maketitle 

A promising strategy to encode information in molecular units
is provided by single molecule magnets (SMMs) \cite{Sessoli03ACIE},
chemically identical nanoscale clusters of exchange-coupled transition
metal or rare earth ions and associated ligands. SMMs have been used
to study quantum tunneling of magnetization and topological quantum
phase interference \cite{Wernsdorfer99S} and may find applications in
quantum information processing
\cite{Tejada01N,Ardavan07PRL,Leuenberger01N}.  The assembly of these
systems on surfaces is currently investigated
\cite{Cornia03ACIE,Condorelli04ACIE,Mannini09NM,Naitabdi05AM,Gomez-Segura06CC,Coronado05IC,Burgert07JACS,Fleury05CC}
as this represents a necessary prerequisite for magnetic memory
applications. Recently, synchrotron-based techniques have been used to
confirm that \Fe\ SMMs remain intact when grafted on a gold surface
preserving their unique magnetic properties
\cite{Mannini09NM}. However, the effect of the surface on an SMM is
still not well understood. This is due to the small quantity of
magnetic material contained in a (sub)monolayer which prevents the use
of techniques such as SQUID magnetometry or conventional nuclear
magnetic resonance (NMR). In this paper we overcome this obstacle by
using a novel proximal magnetometry technique utilizing muons as an
implanted local probe to investigate magnetic properties of a
monolayer of \Fe\ molecules when they are grafted on a Au (111)
substrate. We anticipate that this method will provide a powerful tool
with a high sensitivity over a wide range of time scales and energies,
thus improving our understanding of the influence of the surface on a
grafted SMM.

\begin{figure}[h]
 \centerline{\includegraphics[width=1\columnwidth]{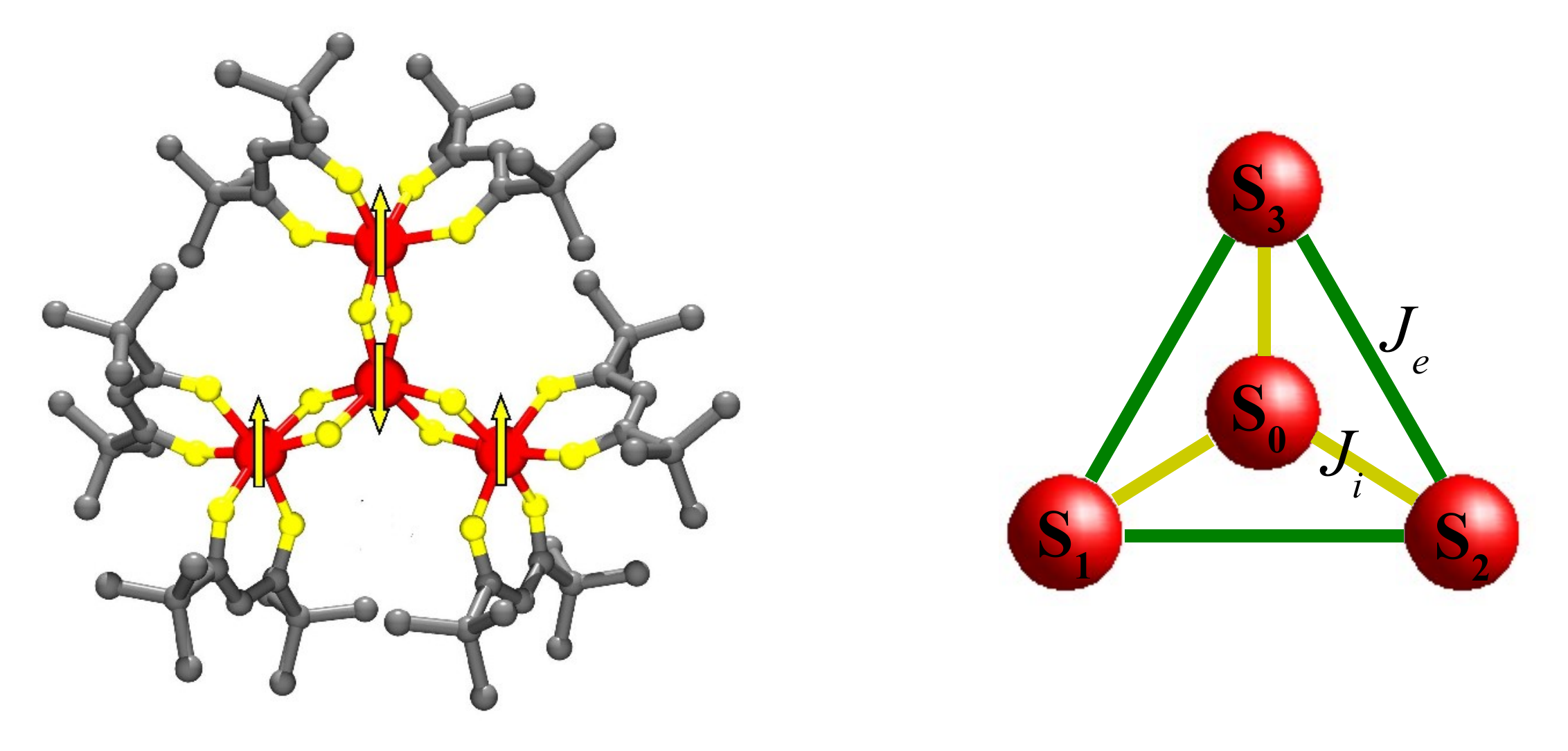}} 
\vspace{-.4cm}
\caption{(color online) A schematic view of the magnetic core of the \Fe\ single
molecule magnet. The red balls denote Fe$^{3+}$ ions while oxygen
atoms are shown in yellow and carbon atoms in gray. On the right panel
we show the labeling scheme of exchange interactions.}
\label{Fe4Core} 
\end{figure}

To date, \Fe\ (Fig.~\ref{Fe4Core}) is the only SMM complex that has
been clearly shown to maintain its SMM behaviour when grafted on a
surface\cite{Mannini09NM}. In the bulk it has a $S=5$ ground state and
a reversal anisotropy barrier up to 17 K\cite{Gregoli09CEJ}. In this
work the properties of the monolayer measured using low energy muon
spin rotation (\lem) method (see below) are compared to conventional
muon spin rotation (\musr) measurements in bulk \Fe\ powder samples.
In both monolayer and bulk, the strength and distribution of the
magnetic dipolar fields from the \Fe\ moments determine the muon spin
precession and relaxation. Interestingly, the qualitative temperature
dependence of the relaxation in both monolayer and bulk are similar,
however, the temperature scales differ. These results are in agreement
with previous X-ray absorption measurements (XAS) \cite{Mannini09NM},
namely they show that the general SMM nature of \Fe\ is preserved in
the monolayer, in contrast with what is observed for the archetypal
SMM Mn$_{12}$ \cite{Salman07NL,Mannini08CEJ}.  Nevertheless, we detect
modifications of the microscopic properties of SMMs compared to bulk
that we attribute to an enhancement of the exchange interaction in the
tetranuclear cluster. \lem\ is thus confirmed as an exceptionally
powerful tool for the investigation of magnetic nanostructures,
providing a complementary view of their magnetic properties.

\begin{figure*}[htb]
 \includegraphics[height=0.8\columnwidth,clip,trim=0mm 0mm 0mm 13mm]{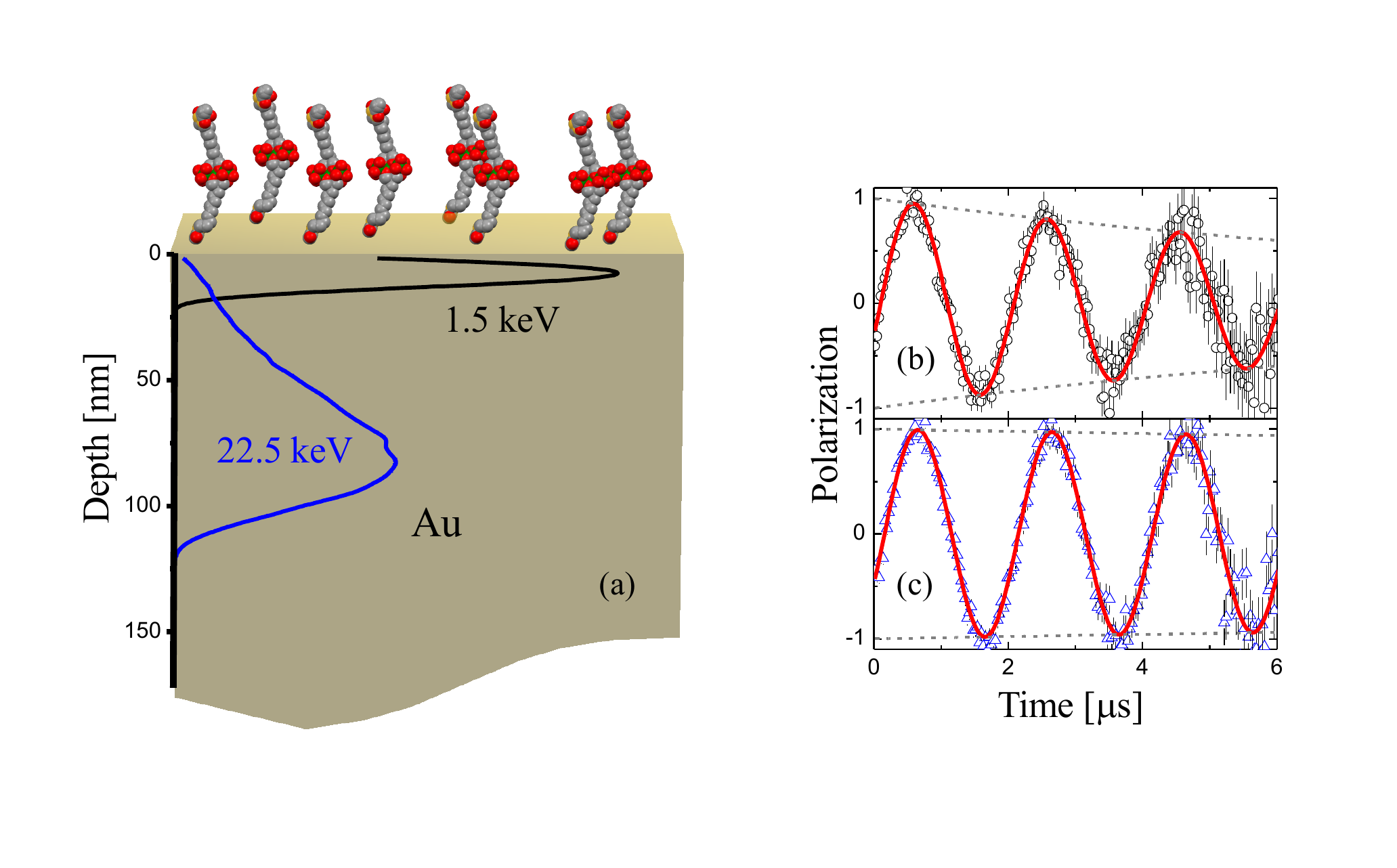} 
\vspace{-1.5cm}
\caption{(color online) (a) A schematic view of sample \textbf{1} where the \FeML\
molecules are grafted on the Au film. The stopping profiles of muons
in Au at $E=1.5$ and $22.5$~keV are also shown. Typical muon spin
precession spectra in the rotating reference frame measured in sample
\textbf{1} at $T=6$ K, $B_{0}=110$ mT and implantation energy (b)
$1.5$ keV and (c) $22.5$ keV. The solid lines are fits to a damping
precession signal and the dashed lines represent the damping
amplitude.}
\label{Asy} 
\end{figure*}
The \lem\ experiments\cite{Morenzoni94PRL,Prokscha08NIMA} were
performed on the low energy muons beamline at PSI in Switzerland. In
these experiments $100\%$ polarized positive muons are implanted in
the sample. The studied samples were mounted in an ultra high vacuum
(UHV) environment on a cold finger cryostat. The experiments described
here were performed in the transverse field (TF) configuration, where
the muon polarization is transverse to the direction of the beam and
the applied magnetic field. Each implanted muon decays (lifetime
$\tau=2.2$ $\mu$s) emitting a positron preferentially in the direction
of its polarization at the time of decay. Using appropriately
positioned detectors, one measures the asymmetry of the muon beta
decay along a certain direction as a function of time $A(t)$, which is
proportional to the time evolution of the muon spin polarization,
$P(t)$, along that direction. $A(t)$ depends on the distribution of
internal magnetic fields and their temporal fluctuations. In \lem\
experiments, the muons implantation energy can be varied between
$1-32$~keV, corresponding to an average implantation depth of
$5-200$~nm, allowing depth resolved \lem\ measurements.

\lem\ measurements on two samples are reported here; sample \textbf{1}
is a monolayer of a \Fe\ (hereafter referred to as \FeML) grafted on a
200 nm thick Au(111) film through
sulfur-functionalization\cite{Gregoli09CEJ,Barra07EJIC}. Sample
\textbf{2} is an identically prepared bare Au(111) film, used as a
control sample in order to confirm that the effects measured in
\textbf{1} are solely due to the grafted \FeML. The functionalized
\Fe\ cluster {[}Fe$_{4}$(L$^{\text{A}}$)$_{2}$(dpm)$_{6}$], where
Hdpm=dipivaloylmethane, and
H$_{3}$L$^{\text{A}}$=11-(acetylthio)-2,2-bis(hydroxymethyl)undecan-1-ol,
synthesized according to Ref.~\cite{Barra07EJIC} has been used to
prepare a monolayer on gold by a wet chemistry approach. This is based
on a three-step process as described in Ref.~\cite{Mannini09NM}; 1)
Hydrogen flame annealing of a $\sim200$ nm thick Au film (evaporated
on mica) to obtain a flat Au(111) surface, 2) incubating the annealed
Au film in a solution of {[}Fe$_{4}$(L$^{\text{A}}$)$_{2}$(dpm)$_{6}$]
in CH$_{2}$Cl$_{2}$ for 20 hours to insure full coverage of the Au
surface, and 3) repeated washing with CH$_{2}$Cl$_{2}$ to remove the
excess of physisorbed molecules and drying the sample in nitrogen
flow. 

The temperature and magnetic field dependence of the \lem\ precession
signals in both samples were measured by implanting the muons at
different energies in the Au film, {\em below} the \FeML\ monolayer.
The magnetic field was applied perpendicular to the Au surface. These
results are compared here to the bulk \musr\ measurements performed on
sample \textbf{3} \cite{SalmanFe4}; a powder sample of an
unfunctionalized {[}Fe$_{4}$(L$^{\text{B}}$)$_{2}$(dpm)$_{6}$]
cluster, where H$_{\text{3}}$L$^{\text{B}}$ is the commercially
available tripodal ligand 1,1,1-tris(hydroxymethyl)ethane, prepared
according to Ref.~\cite{Cornia04ACIE} (hereafter referred to as \FeB).
In the bulk, both functionalized and unfunctionalized \Fe\ clusters
have very similar spin Hamiltonian parameters, in particular the
nearest-neighbor exchange interaction between Fe spins are $J_{i}=
22.2-22.9$~K (depending on the crystal phase)
\cite{Gregoli09CEJ,Barra07EJIC} and 23.8 K \cite{Cornia04ACIE} for the
functionalized and unfunctionalized clusters, respectively.

Typical \lem\ precession signals recorded at $T=6$ K and two different
implantation energies ($E$) are shown in Fig.~\ref{Asy}. The
corresponding muon stopping profiles at 1.5 and 22.5 keV implantation
energies are shown in Fig.~\ref{Asy}(a). The muon polarization follows
a precessing damping signal
\begin{equation} P(t)=P_{0}\cos(\omega
  t)e^{-\lambda t},\label{Pol}
\end{equation}
where $P_{0}$ is the initial polarization, $\omega$ is the precession
frequency and $\lambda$ is the damping/relaxation rate of the
oscillations.  In these measurements, the precession frequency is
$\omega=\gamma B$, where $\gamma=135.5$ MHz/T is the muon gyromagnetic
ratio and $B$ is the average local magnetic field experienced by the
muons. The relaxation rate $\lambda$ gives a measure of the width of
the distribution of local magnetic fields, thus a wider field
distribution results in a larger $\lambda$. Note that this parameter
is primarily a measure of the local static magnetic fields, but may
also include small contributions from dynamic magnetic fields as well.
In general, muons are sensitive to magnetic field fluctuations in the
range $10^{-11}-10^{-4}$ s. Faster fluctuations are averaged out while
slower fluctuations are static over the lifetime of the muon (2.2
$\mu$s).

\begin{figure}[h]
 \centerline{\includegraphics[width=1\columnwidth]{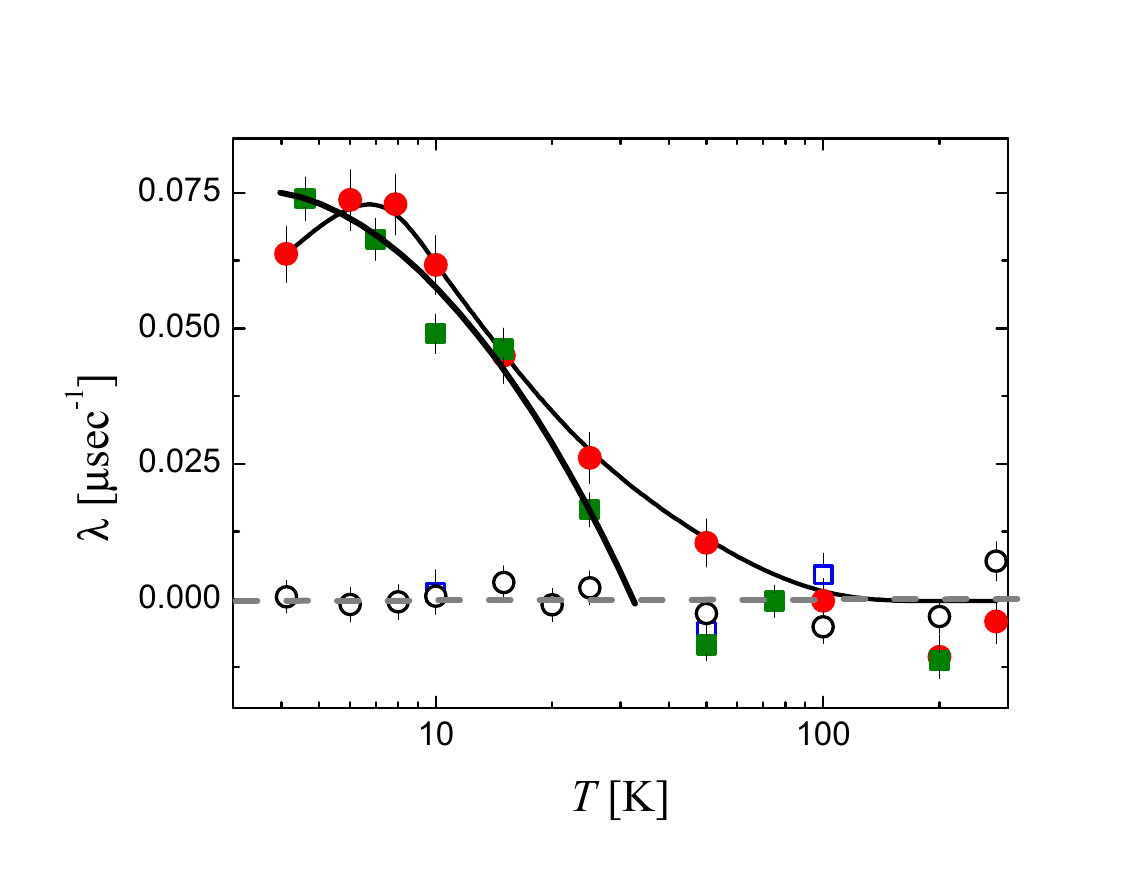}} 
\vspace{-.7cm}
\caption{(color online) The relative relaxation rate measured in
  sample \textbf{1} (full symbols) and \textbf{2} (open symbols). The
  squares and circles indicate measurements in $B_{0}=8.3$ mT and 110
  mT, respectively. The lines are a guide to the eye.}
\label{ML} 
\end{figure}
The average local field measured at all temperatures and implantation
energies (in both samples \textbf{1} and \textbf{2}) was equal to the
applied field, $B_{0}$. However, the distribution of fields measured
by $\lambda$ exhibits a strong $T$ and $E$ dependence only in sample
\textbf{1} but not in \textbf{2}. For example, in Fig.~\ref{Asy}(b) we
plot $P(t)$ at $E=1.5$~keV, where most of the muons stop within
$10-20$~nm of the Au surface, and the dipolar fields from the \FeML\
moments are large \cite{Salman07NL}. In contrast, in Fig.~\ref{Asy}(c)
we plot $P(t)$ at $E=22.5$~keV where the average muon implantation
depth is $\sim100$~nm, and the dipolar fields at this depth are
negligible. Therefore, the measured $\lambda$ at $1.5$~keV is larger
than that at $22.5$~keV. Note that the dipolar fields sensed by the
implanted muons are proportional to the average size of the \FeML\
magnetic moment, which is determined by the population of the
different spin states of \FeML \cite{Gregoli09CEJ,Barra07EJIC}.

The strong temperature dependence of $\lambda$ measured at low energy
is shown in Fig.~\ref{ML}. Here we plot $\lambda$ in sample \textbf{1}
relative to \textbf{2}; i.e. zero relaxation is taken as the average
relaxation measured in the reference sample (since there is no
temperature and field dependence). Note that the precession signals
measured in sample \textbf{2} are identical to those measured at high
energy in sample \textbf{1} and do not depend on $T$ or $E$. %

At high temperatures ($T\gg J_{i}$) and $E=1.5$ keV the relaxation
rate is temperature independent and small (similar to sample
\textbf{2}). This is due to fast thermal fluctuations of the
individual Fe magnetic moments, which average out the dipolar magnetic
field experienced by muons and therefore produces a narrower field
distribution (smaller $\lambda$). Below $\sim40-50$ K the relaxation
rate increases rapidly with decreasing temperature, indicating a
slowing down of the fluctuations of the Fe moments and an increase in
the magnetic correlations within each molecule. Therefore, the average
magnetic moment per molecule and hence the dipolar fields on the muons
increase. At $B_{0}=110$ mT the relaxation rate peaks at $\sim6$ K and
then decreases upon further cooling, in contrast to the monotonic
increase at $B_{0}=8.3$ mT.  The difference between the temperature
dependencies at different fields is due to the strong field dependence
of the \Fe\ spin energy levels and its magnetic moment
\cite{Gregoli09CEJ}.

\begin{figure}[h]
\centerline{\includegraphics[width=1\columnwidth]{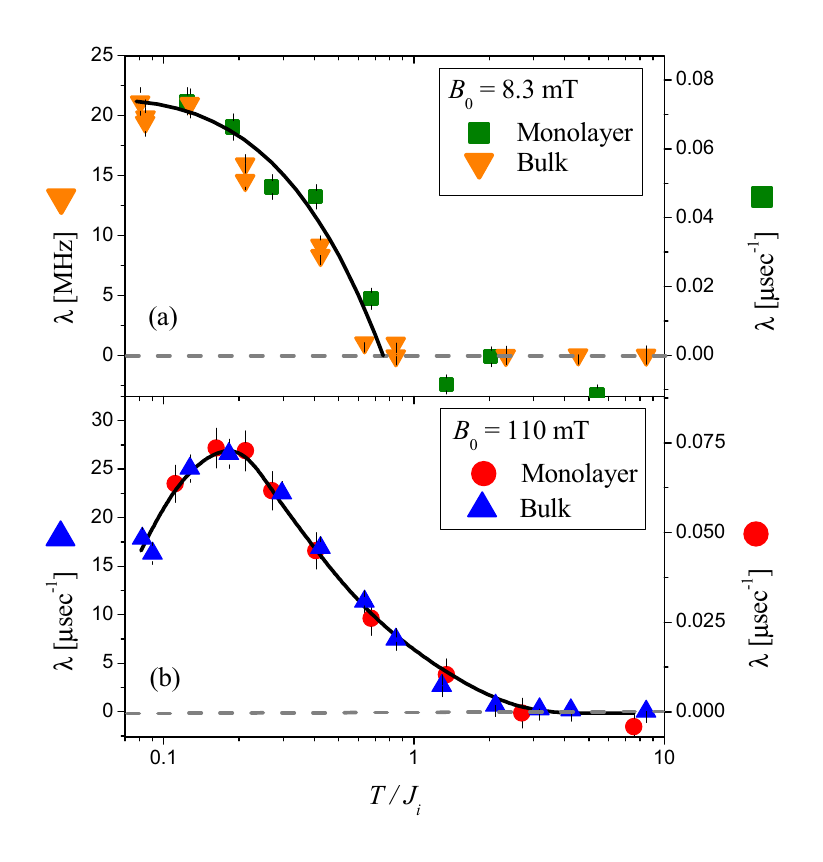}} 
\vspace{-.7cm}
\caption{(color online) The relaxation rate as a function of the
  normalized temperature measured in bulk \Fe\ compared to monolayer
  at fields (a) $8.3$ and (b) $110$ mT. The lines are a guide to the
  eye.}
\label{MLvsBulk} 
\end{figure}
Similar bulk \musr\ measurements were done on a powder \FeB\ sample,
in the same temperature range and applied field transverse to the muon
spin direction. In these measurements, we detect two precession
signals due to two inequivalent muon sites. The temperature dependence
of the precession parameters are qualitatively similar for both sites,
and therefore we compare the measurements in the monolayer to the
precession signal of the majority of the muons in bulk. The relaxation
rates measured at $B_{0}=8.3$ and $110$ mT are shown in
Fig.~\ref{MLvsBulk}(a) and (b), respectively, along with the
corresponding data in \FeML\ . As in the case of the \lem\
measurements, these relaxation rates are due to the distribution of
dipolar fields in bulk at the respective applied fields. Therefore,
the underlying mechanism of spin relaxation in both cases is
identical.

The observed temperature dependence of the relaxation rates is typical
of molecular nanomagnets and reflects the building up of exchange
correlations among the spins of the molecule, widely investigated
using NMR and \musr\ \cite{Borsa07,Salman02PRB,Salman08PRB}. As
expected, the temperature dependencies measured in the monolayer have
striking qualitative resemblance to those measured in bulk. However,
the ratio of the relaxations in \FeB:\FeML\ is $\sim300:1$ indicating
that the muons in the monolayer are on average $\sim7$ times further
from the \Fe\ moments compared to bulk, and therefore experience much
smaller dipolar fields. Interestingly, the temperature dependence in
the monolayer seems to be shifted to higher temperatures compared to
bulk. This shift is almost logarithmic in temperature, indicating that
it occurs due to a temperature independent increase in the energy
scales of \FeML\ compared to \FeB.

Given the sensitivity of muons to the energy spectrum arising from
exchange interactions in magnetic molecules
\cite{Borsa07,Salman02PRB,Salman08PRB} we attempt to rationalize the
observed shift by using the following spin Hamiltonian for \Fe\
\cite{Barra07EJIC}
\begin{equation}
{\cal H}_{0}=J_{i}{\mathbf{S}_{0}}\cdot\sum_{k=1}^{3}{\mathbf{S}_{k}}+J_{e}\left[{\mathbf{S}_{1}}\cdot{\mathbf{S}_{2}}+{\mathbf{S}_{2}}\cdot{\mathbf{S}_{3}}+{\mathbf{S}_{3}}\cdot{\mathbf{S}_{1}}\right],
\end{equation}
where $J_{i}$ is the super-exchange coupling between the central
$({\mathbf{S}_{0}})$ and the three peripheral $({\mathbf{S}_{k}})$ Fe
ions, and $J_{e}$ is the next nearest neighbour interaction
(Fig.~\ref{Fe4Core}). Since $J_{i} \gg J_{e}$ the landscape of total
spin levels is determined mainly by $J_{i}$. In a simplified picture
we interpret the temperature dependence of $\lambda$ as follows.  For
$T\gg J_{i}$ the individual Fe magnetic moments are thermally
fluctuating. These fluctuations slow down as $T$ approaches $J_{i}$,
and continue to do so at lower temperatures as correlations between
the individual Fe moments within each molecule are formed. This is
also accompanied by an increase in the average magnetic moment of the
molecule. The increase observed in the relaxation rate is associated
with the increase in dipolar fields due to the larger moments as well
as the slowing down of the fluctuations
\cite{Salman02PRB,Salman08PRB,Lascialfari98PRL}.  Since $J_{i}$
provides the basic energy scale in this system (at high temperatures),
it is intuitive to plot the relaxation rate as a function of the
normalized temperature, $T/J_{i}$ (Fig.~\ref{MLvsBulk}).  For the \Fe\
clusters used here $J_{i}^{B}$ evaluated from magnetic measurements is
found to be 23 (1) K \cite{Cornia04ACIE,Gregoli09CEJ,Barra07EJIC}, and
therefore the temperature dependence of the relaxation rate in \FeB\
shows a dramatic increase below $T/J_{i}^{B}\sim1$. In contrast, the
value of the coupling $J_{i}^{ML}$ in \FeML\ is unknown. However, in
order to obtain a similar behaviour to that seen in bulk, one has to
plot the relaxation rate in the monolayer against a normalized
temperature $T/J_{i}^{ML}$, where $J_{i}^{ML}=37(1)$ K. This clearly
produces an excellent agreement between the relaxation rates measured
in bulk and monolayer for all temperatures and magnetic fields,
suggesting that the main difference between \Fe\ in bulk and monolayer
can be accounted for by a $\sim60\%$ increase in $J_{i}$ of \FeML\
over \FeB. Enhancement in the Fe-Fe $J$ coupling as a function of the
Fe-O-Fe angle has been observed in iron dimers \cite{LeGall97ICA}, and
predicted using first principle density functional theory calculations
\cite{Stolbov05condmat}. We should also take into account that thermal
fluctuations of the spins are governed by the spin-phonon coupling and
sound velocity \cite{GatteschiMNN}. These are also expected to be
dramatically altered by the different environment of the monolayer,
and could result in the apparent change in energy scale. However, it
is less likely that this effect plays a dominant role since it would
not explain the observed scaling between monolayer and bulk.

In conclusion, our results have demonstrated that the general SMM
nature of \Fe\ is preserved in the monolayer, in contrast to the case
of Mn$_{12}$\cite{Salman07NL,Mannini08CEJ}. The observation of a
different temperature scale of the magnetic properties of the
monolayer suggests an enhancement of the intramolecular interactions
between the individual Fe ions within the \Fe\ core. This implies a
non negligible role played by the surface, which, once understood,
could be exploited to tune the magnetic properties of nanostructured
SMMs. By using polarized muons as proximal magnetometers, we have
developed a very powerful technique able to characterize the magnetic
properties of exceedingly small amounts of magnetic molecules, over a
wide range of time scales and energies. This capability opens the
possibility to investigate the details of the effects of surfaces on
the magnetic properties of molecular nanostructures, which is not
possible using other conventional techniques.

This work was performed at the Swiss Muon Source S$\mu$S, Paul
Scherrer Institute, Villigen, Switzerland. We would like to thank
Hans-Peter Weber for technical assistance and Giovanni Aloisi for the
help in the preparation of gold substrates. The financial support of
the EU through the NoE MAGMANet , and through NANOsci-ERA and
MOLSPINQUIP projects is gratefully acknowledged.

\newcommand{\noopsort}[1]{} \newcommand{\printfirst}[2]{#1}
  \newcommand{\singleletter}[1]{#1} \newcommand{\switchargs}[2]{#2#1}

\end{document}